\documentclass[12pt]{article}
\usepackage{authblk}
\usepackage[T1]{fontenc}
\usepackage[english]{babel}
\usepackage[latin9]{inputenc}
\usepackage{amstext}
\usepackage{amssymb,amsmath}
\usepackage{graphicx}
\usepackage{babel}

\usepackage{graphics}
\def\beq{\begin{equation}}
\def\eeq{\end{equation}}

\providecommand{\keywords}[1]
{
  \small	
  \textbf{\textit{Keywords---}} #1
}

\begin{document}
\title{Critical velocity in resonantly driven  polariton  superfluids}

\author{Simon Pigeon} \affil{Laboratoire Kastler Brossel, Sorbonne Universit\'e, CNRS, ENS-PSL Research University, Coll\`ege de France, 4 place Jussieu, 75252 Paris, France}
\author{Amandine Aftalion} \affil{\'Ecole des Hautes \'Etudes en Sciences Sociales,  Centre d'Analyse et de Math\'ematique Sociales, UMR-8557, 54 boulevard Raspail, Paris, France.}

\date{\today}
\maketitle

\begin{abstract}
We study the necessary condition under which a resonantly driven exciton polariton
superfluid flowing against an obstacle can generate turbulence. The value of the critical velocity is well estimated by the transition from elliptic to hyperbolic of an operator following ideas developed by Frisch, Pomeau, Rica \cite{FPR} for a superfluid flow around an obstacle, though the nature of equations governing the polariton superfluid is quite different. We find analytical estimates depending on the pump amplitude and on the pump energy detuning, quite consistent with our numerical computations.
\end{abstract}

\section{Introduction}

Since the discovery of Helium in 1937, superfluidity has attracted some of the greatest minds of our time. After nearly one century of studies, this phenomenon does not stop puzzling our understanding of matter. First observed in liquid Helium \cite{allen1938flow,kapitza1938viscosity}, superfluidity has been studied in details more recently in atomic condensates \cite{anderson1995observation,O,raman99,RO}. In analogy with the rotating bucket experiment in Helium \cite{don}, quantized vortices have been observed in rotating one component \cite{mad,AK} and two component condensates \cite{mat}. Superfluid physics has now spread far beyond the field of atomic physics and is used to describe the behavior of a large variety of system, from non-linear optical system \cite{Nice,Rb} to neutron star \cite{MIGDAL1959655}  or bird flocks \cite{Attanasi_2014}.

In this paper, we address the issue of the existence of a dissipationless flow induced
 by the motion of a macroscopic object in a superfluid. The nucleation of vortices corresponds to the
  breakdown of this dissipationless phenomenon. A classical experiment on superfluid Helium consists in flowing
 Helium around an obstacle. If the velocity of the flow at infinity is sufficiently small,
 the flow is stationary and dissipationless, as opposed to what happens in a normal fluid.
 On the other hand, beyond a critical velocity, the flow becomes time dependent and vortices are emitted
 periodically from the north and south pole of the obstacle. Numerical simulations illustrating
 this behaviour have been performed by Frisch, Pomeau, Rica \cite{FPR}: a pair of
 vortices is emitted and is flowing behind the obstacle, while the next pair
 is being formed on the boundary of the obstacle. In Ref. \cite{FPR}, the authors have also computed the critical
 velocity for the nucleation of vortices. Other related works, that we
 will describe below, include \cite{hakim,HB,JMA,rica,JPR}. The absence of dissipation at low velocity can be explained by the existence of a stationary solution to some two-dimensional nonlinear Schr{\"o}dinger equation.
 The superfluid velocity is given at any point in the flow by the
  gradient of the phase of the wave function: if the wave function
 does not vanish, then the velocity is well defined everywhere.
 The vortices are points where the wave function vanishes and
 around which the circulation of the velocity is quantized.

Following these theoretical works, an experiment was conducted at MIT by Raman et al.
\cite{raman99}, (see also \cite{O,RO}) in Bose
 Einstein condensates, to study there the existence of a dissipationless
 flow. Instead of a macroscopic object, the
 obstacle is a blue detuned laser beam. The condensate is
 fixed and the obstacle is stirred in the condensate. Similar
  features to Helium are observed, namely the evidence of a
critical velocity for the onset of dissipation. The energy release
is measured as a function of the velocity of the stirrer: if the
velocity is small, the flow is almost dissipationless and the drag
on the obstacle is very small, while above a critical value of the
velocity, the flow becomes dissipative. Numerical simulations have
been performed by \cite{ADP} for the 3D problem corresponding to the experiment, relating the increase in energy
dissipation to vortex nucleation.

Among the systems where superfluidity is observed, exciton-polariton fluids have attracted significant attention as to their ease of control and manipulation thanks to their dual light-matter nature. Exciton-polariton fluids are composite bosons resulting from the strong coupling between the excitonic resonance of a semiconductor quantum well and the microcavity  electromagnetic field \cite{carusotto2013quantum}. In particular, the experimental study of a polariton field flowing past an obstacle, and the observation of quantized vortices in the wake of the obstacle has been the subject of many papers \cite{amo2009superfluidity,amo2011polariton,Nardin2011}. This superfluid and turbulent behaviours have been the topic of quite a few theoretical papers \cite{PCC,jug,PA,amecar} and this is at the core of our study.
   Mixing advantageously the low effective mass of cavity light with the strong inter-particle interaction between matter excitation such as semiconductor excitons, these systems have shown recently superfluid behavior even until room temperature \cite{lerario2017room}.

In many circumstances such as the one considered here, it is not necessary to work with the pair of equations of motions for the photonic
and excitonic fields and one can restrict to a single classical field describing the lower polariton field. This simplified description is generally legitimate provided the Rabi frequency is much larger than all other energy scales of the problem, namely the kinetic and interaction energies, the pump detuning, and the loss rates $\gamma$ \cite{carusotto2013quantum}. Contrary to atomic superfluids, polariton superfluids are driven-dissipative fluids. To compensate their short lifetime (of the order of tens of picoseconds), the system must be continuously pumped. Here, we consider continuous and quasi-resonant pumping of frequency $ \omega_p$
 and amplitude $F$. One of the interests of this technique to create a polariton fluid, is that it allows the creation of a flowing fluid. If the laser beam is slightly tilted with respect to the cavity plan, the polariton fluid generated within the plan of the cavity will carry a finite momentum $\mathbf{k}_p$. Consequently, in a polariton fluid, contrary to what happens in a cold-atomic ensemble, the obstacle is fixed and the fluid is moving at a speed far from the obstacle which is $v_\infty=\hbar |\mathbf{k}_p|/m$.
  This yields the following generalized Gross-Pitaevskii equation (GPE) or Nonlinear Schr\"odinger equation (NLS) for the polariton field $\psi$ in the pump rotating frame:
\begin{equation}
i\hbar \partial_{t}\psi(\mathbf{x},t)= \left(-\frac{\hbar^2}{2m}\nabla^{2}-\Delta-i\frac{\gamma}{2}+V(\mathbf{x})+
g\vert\psi(\mathbf{x},t)\vert^{2}\right)\psi(\mathbf{x},t)+Fe^{i\mathbf{k}_{p}.\mathbf{x}}\label{eq:GP}
\end{equation}
 where  $m$ is the polariton effective mass set to 1 and $g$ is the interaction strength. Contrary to atomic superfluids, no confinement potential is needed. The other parameters are directly linked to the driven-dissipative nature of polaritons: $\gamma$ is the decay rate of polaritons, $\Delta$ the energy detuning between the driving field $\hbar\omega_p$ and the polariton eigenenergy, $F$ the coherent driving field and $\mathbf{k}_{p}$ the driving field momentum. The potential $V(\mathbf{x})$ is an added repulsive potential modelling the obstacle, which is therefore equal to 0 outside the obstacle.

 When varying the driving field $F$, the polariton density undergoes an S-like dependency with a bistable regime, presenting a low-density regime and a high-density regime \cite{amo2009superfluidity,carusotto2013quantum}. In the low-density regime the interaction term in Equation \eqref{eq:GP} can be neglected to lead to a standard linear system. On the contrary, in the high-density regime, the interaction term cannot be neglected and leads to the appearance of a superfluid behavior.
  In the first experiments, it was thought that the driving  inhibits the formation of vortices. Therefore, in order to observe the nucleation of vortices past an obstacle,
  the fluid was released from the driving presence either temporally \cite{PCC,Nardin2011} or spatially \cite{amo2011polariton}. A detailed numerical studied recently revealed a more subtle situation \cite{PA}: indeed, the driving field tends to inhibit the formation of turbulence, however, it can be reduced enough to release its constraints and allows  the formation of vortices on the edge of the obstacle. Fine-tuning of the driving amplitude $F$ eventually allows passing from a dissipationless superfluid to a turbulent one \cite{PA} without having to remove it. This has been achieved experimentally very recently \cite{Ler}.




\begin{figure}
\begin{centering}
\includegraphics[width=.7\columnwidth]{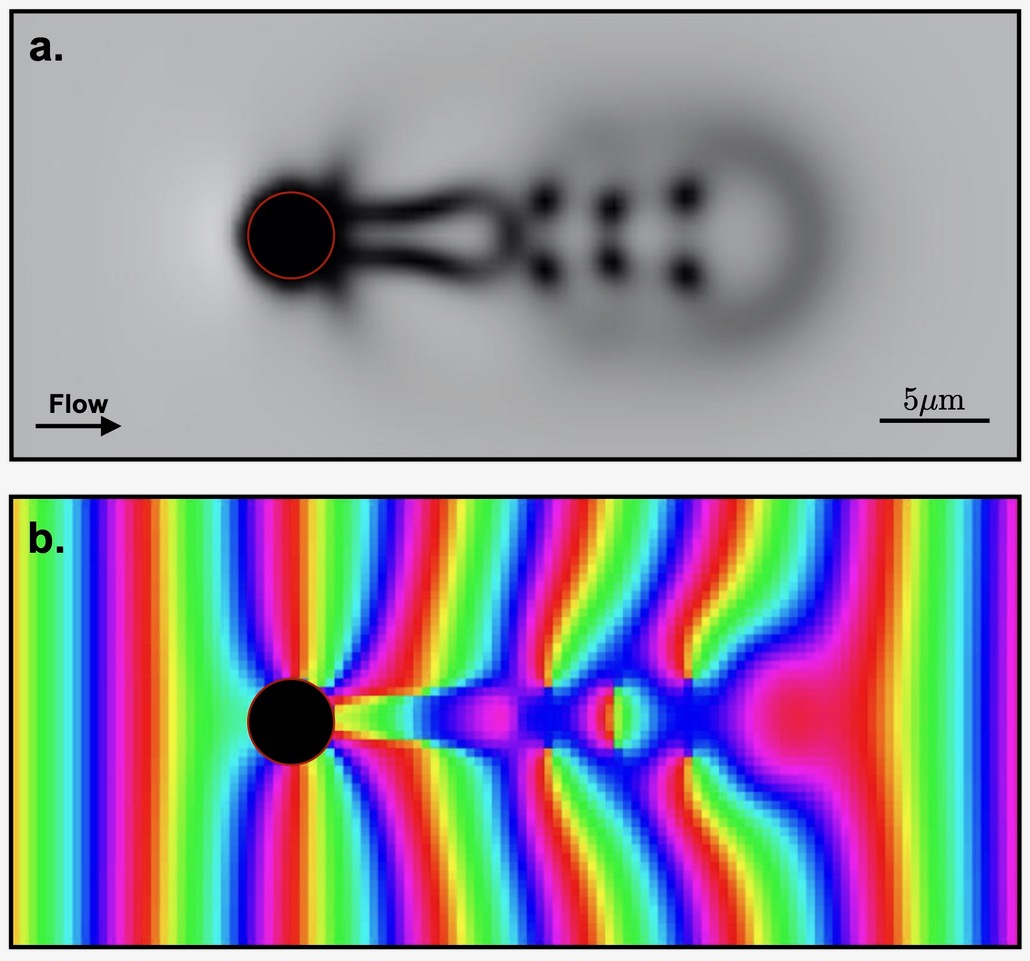}
\par\end{centering}
\caption{Polariton fluid density in panel a., while propagating against an obstacle marked by a red circle leading to the appearance of vortex in its shadow. Corresponding phase profile in panel b. with the obstacle region marked in black.
Physical parameters:
 $k_{p}=0.718 \:\mu\text{m}^{-1}$, $m=1$~meV.ps$^2.\mu\text{m}^{-2}$, and  $  F=0.24$ meV, $\Delta=0.7$ meV, $ \gamma=0.05$ meV, $  g=0.01$ meV;
 simulation parameters: $256\times128$ grid, $dt=0.02$ ps.}
\label{fig:intro}
\end{figure}

Whereas in atomic superfluids, the Mach number $M=v_\infty/c_s$ (where $c_s=\sqrt{g |\psi|^2/m}$ is the fluid speed of sound), is the only parameter controlling the transition from dissipating energy via vortex emission to dissipationless, in the present driven-dissipative scenario, the pump field amplitude plays a crucial role. In this work, we will focus on this phenomenon and disentangle the role played by the pump amplitude $F$ and pump detuning $\Delta$ in this transition from turbulent to a non-turbulent superfluid.

{ We will perform numerical simulations of Equation \eqref{eq:GP} such as in Figure \ref{fig:intro} which illustrates the vortex nucleation behaviour and will be detailed below. But we will also perform an analytical approach of the transition behaviour.}

\section{Equations}

 As mentioned above, this problem of onset of dissipation in a superfluid was first addressed by
 Frisch, Pomeau and Rica \cite{FPR}. They have studied the case
where the obstacle is a small disk in the frame
where the obstacle is fixed.
 The nonlinear Schr\"odinger equation studied in \cite{FPR} can be rewritten using the hydrodynamic formulation, where $\rho$ is the density and
 which allows to identify $\nabla\phi$ with a velocity: \beq
 {\frac{\partial \rho}{\partial t}}+\nabla \cdot (\rho\nabla
\phi)=0\label{cons}\eeq
\beq{\frac{\partial \phi}{\partial
t}=\frac{\Delta\sqrt\rho}{2\sqrt\rho}-\rho+c_s^2+\frac 12 v_\infty^2-\frac 12|\nabla\phi|^2},\label{cons2}
\eeq   where the mass $m$ is set to unity, $c_s$ is the sound speed, $v_\infty$ the flow velocity at infinity.
 They look for stationary solutions and assume
that the quantum pressure term $\Delta \rho/\sqrt{\rho}$ is
negligible, which is a kind of long wave approximation, and
leads to the following problem \beq\label{diveq} \nabla\cdot (\rho\nabla
\phi)=0,\quad \rho=c_s^2+\frac 12 v_\infty^2-\frac 12|\nabla\phi|^2,\eeq with boundary
conditions $\partial \phi/\partial n =0$ on the obstacle,
$\rho\to c_s^2$,  and $\nabla \phi\to v_\infty$ at infinity. Note that the
second equation in (\ref{diveq}) is a Bernouilli law for this
problem. The system (\ref{diveq}) has the same
formulation as
 that of a stationary
irrotational flow of a compressible fluid about an obstacle.
 Mathematically, the
  existence of solutions for such a related subsonic problem  and the non existence for large velocity at infinity
  is proved by
\cite{toto,Sc} using a fixed point theorem. In one dimension, in Ref. \cite{hakim}, the  saddle node bifurcation is analyzed  and the critical velocity is related to the spatial variation of the potential representing the obstacle.

Let $v=\nabla \phi$. From \cite{FPR}, Equation \eqref{diveq} goes from elliptic to hyperbolic when
\begin{equation}
\partial_{v}(v\rho(v))=0, \hbox{ with } \rho(v)=c_s^2+\frac 12 v_\infty^2-\frac 12 v^2.\label{eq:criteria_reduce}
\end{equation}This yields that the transition first takes place when \begin{equation}c_s^2+\frac 12 v_\infty^2-\frac 32 v^2=0.\label{vcrit}\end{equation}
 As explained in Ref. \cite{FPR}, this first happens at the point of maximum of $v$, which is at the north and south pole of the obstacle. It is therefore crucial to estimate this maximal velocity. Several papers deal with this question \cite{HB,W,rica}.  The maximal velocity occurs at the equator of an object and is $3/2 v_\infty$ for a sphere and $2 v_\infty$ for a cylinder. Nevertheless for a compressible fluid, this equatorial velocity is slightly larger due to pressure effects. In Ref. \cite{rica}, an asymptotic expansion in terms of the Mach number is made.
 Further studies about this critical velocity in particular including scaling laws and the bifurcation diagram can be found in Ref. \cite{HB}. What happens beyond this critical velocity and the transition to an Euler-Tricomi equation has been studied in Ref. \cite{JPR}.
 In Ref. \cite{FPR}, it is assumed that the maximal velocity is $2 v_\infty$ which yields from  Equation \eqref{vcrit}
 $$v_\infty=\sqrt{\frac 2 {11}}c_s$$ for the onset of dissipation.

 Similarly to Ref.  \cite{FPR},  in our case of a polariton superfluid, Equation \eqref{eq:GP} can be rewritten in terms of the phase and amplitude using the Madelung transform so that $\psi=\sqrt{n}e^{i\theta}$ which yields, after setting $m$ to 1,
\begin{align}
\hbar \partial_{t}n & =-\hbar^2\nabla.\left(n\nabla\theta\right)-2n\left(\frac{\gamma}{2}-\frac{F}{\sqrt{n}}\sin(\mathbf{k}_{p}.\mathbf{x}-\theta)\right)
\label{eq:cont}\\
\hbar \partial_{t}\theta & =\frac{\hbar^2}{2}\left(\frac{1}{\sqrt{n}}\nabla^{2}\sqrt{n}-\left(\nabla\theta\right)^{2}\right)+
\Delta-V-gn-\frac{F}{\sqrt{n}}\cos(\mathbf{k}_{p}.\mathbf{x}-\theta)\label{eq:flux}
\end{align}
We assume the flow to be directed along the $x$ axis so that $\mathbf{k}_{p}=(k_{p},0)$ with $k_{p}\ge0$ and $\mathbf{v}=\hbar\nabla\theta$ is the flow speed.
With respect to the Gross-Pitaevskii equation  describing atomic Bose-Einstein condensates \eqref{cons}-\eqref{cons2}, Equations \eqref{eq:cont}-\eqref{eq:flux} include additional terms to account for the driven-dissipative nature of the polariton gas, namely a loss rate proportional to $\gamma$ and the coherent pumping proportional to $F$.

To determine whether the fluid can remain superfluid, we will follow the same approach as in Ref. \cite{FPR}. It consists in determining when the continuity equation, Equation (\ref{eq:cont}), in its stationary version, goes from elliptic to hyperbolic. This only depends on the structure of the operator $\nabla.\left(n\nabla\theta\right)$ and not the right hand side term of the same equation. Therefore, we need  to determine how the density $n$ is connected to the flow velocity $\mathbf{v}=\hbar \nabla \theta$. To do this, we use Equation (\ref{eq:flux}) in its stationary version, away from the obstacle where $V=0$, and we neglect the quantum pressure term. This yields
\begin{equation}
\frac{1}{2}v^{2}+{\frac{F}{\sqrt n}}\cos(k_{p}x-\theta)+gn-\Delta=0.\label{eq:ber}
\end{equation} We point out that this equation is a polynomial of degree 3 in $\sqrt n$.

\section{Results}

We both want to simulate Equation \eqref{eq:GP} and see numerically the change of behaviour on the one hand, and use the hydrodynamic formulation and the change of the operator from elliptic to hyperbolic to find a critical pump amplitude consistent with the numerics.

\subsection{Numerical results}

\begin{figure}
\begin{centering}
\includegraphics[width=.9\columnwidth]{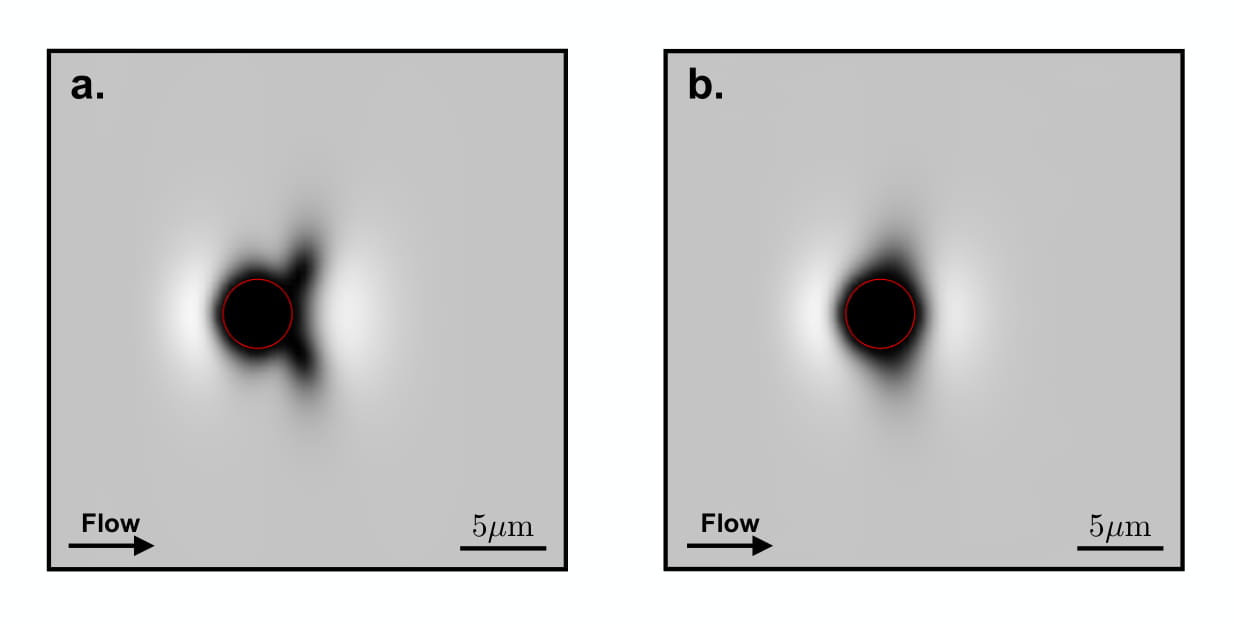}
\par\end{centering}
\caption{Density profile $n/n_{\infty}$ for two different pump
powers: in a. $  F=0.2515$ meV and in b. $ F=0.2712$ meV. The red circle indicate
the position of the potential barrier. The color scale is identical for both images. Physical parameters:   $k_{p}=0.718 \;\mu\text{m}^{-1}$,
$m=1$~meV.ps$^2.\mu\text{m}^{-2}$, $ \Delta=0.3$ meV, $\gamma=0.05$ meV, $  g=0.01$ meV; simulation
parameters: $256\times128$ grid, $dt=0.02$ ps. }
\label{fig:density}
\end{figure}

In the simulations, we fix $k_{p}=0.718 \;\mu\text{m}^{-1}$,
$m=1$~meV.ps$^2.\mu\text{m}^{-2}$,  $\gamma=0.05$ meV, $  g=0.01$ meV. We observe two types of behaviours: for fixed pump detuning $\Delta$, and small pump amplitude $F$, vortices are emitted periodically from the north and south poles as in \cite{FPR} as illustrated in Figure \ref{fig:intro}. On the contrary, for large pump amplitude $F$, the solution is superfluid and there is a stationary flow as illustrated in Figure \ref{fig:density}b.  So for each $\Delta$, there is a critical value of the pump amplitude $F_c$ where the transition goes from a solution emitting vortices to a superfluid solution. Close to the transition, the period of emission gets very large. The difference with Helium or Bose Einstein condensates is that instead of prescribing a velocity at infinity, it is the pump amplitude which is prescribed and characterizes the behaviour.

We simulate Equation \eqref{eq:GP} using a split-step method on a grid made of $256\times128$ pixels corresponding to $100\times50\:\mu\text{m}^2$. We use periodic boundary conditions and tailor the pump profile at the edge of the grid to avoid undesired propagation of density modulation. The time step is $0.2$ ps and we start with an empty system. Then the pump amplitude is slowly raised up to about ten times the pump amplitude of interest. Finally, the amplitude is decreased to reach the elected one, which is then kept constant. This initialization procedure lasting about 200 ps allows us to prepare the system in the upper part of the bistable regime, corresponding to the point where vortex nucleation was numerically observed \cite{PA}.

In Figure \ref{fig:density} we represent the polariton field density around an obstacle (indicated by a red circle) at a given time frame. The only difference between Figures \ref{fig:density}.a and b is the pump amplitude $F$ which varies by less than 10\%. As visible, for low enough pump amplitude a pair of vortex has formed at the edge of the obstacle (\ref{fig:density}.a) whereas for a slightly higher pump amplitude no vortex forms (\ref{fig:density}.b). Both images correspond to the same time frame.

To evaluate numerically the critical pump amplitude $F_c$ (and its dependencies with respect to $\Delta$), for each $\Delta$, we perform a large set of simulation at fixed $F$. For each one, while running we evaluate the formation of vortices at the edge of the defect and stop the simulation if one is found (low-density core distinguishable from the obstacle). We re-run the simulation slightly increasing $F$. If no vortex forms on the edge of the defect after a time long enough (0.5 ns), the simulation is re-run with a slightly lower $F$. The program stops after 8 iterations given accurate enough estimation of the critical pump amplitude $F_c$. Notice that the closer we are to $F_c$ from below, the slower  the nucleation of vortices. Consequently, a significant increase of precision on the determination of $F_c$ will have required an important computational effort not necessary given the result obtained.

It is noticeable that despite the important change of the detuning $\Delta$, the critical driving amplitude $F_c$ only slightly varies around $ 0.26 $ (between .255 and .27).
 Since, $F_c$ is almost constant when $\Delta$ varies (and $k_p$ and $\gamma$ are fixed), this could provide a new relation between the parameters of the problem using the well known relation:
  \begin{equation}\frac{F}{\sqrt{n_\infty}}=\sqrt{\frac{\gamma^2}4+(\frac{v_\infty^2}2-\Delta+gn_\infty)^2}.\label{fn}\end{equation}
 This can be obtained from our system, on the one hand from
 Equation \eqref{eq:ber}, since at infinity, $v_{\infty}=\hbar k_{p}$ the asymptotic fluid speed far
from the obstacle, which is fixed by the driving field momentum $k_{p}$, so we can write $\theta$ as $k_px+\phi$ and we find, from Equation \eqref{eq:ber}, at infinity,
\begin{equation}
\frac{F}{\sqrt{n_\infty}}\cos \phi_\infty=\frac{v_{\infty}^{2}}{2}-\Delta+gn_\infty\label{eq:fc}
\end{equation} and from Equation \eqref{eq:cont}, $$\frac \gamma 2= \frac{F}{\sqrt{n_\infty}} \sin \phi_\infty.$$ Adding the two, we obtain Equation \eqref{fn}.
 We have checked  numerically that Equation \eqref{fn} holds for our solutions. Note that the number of solutions in $n_{\infty}$ to  Equation \eqref{fn} is related to the bistability mentioned in the introduction.

\begin{figure}
\begin{centering}
\includegraphics[width=.75\columnwidth]{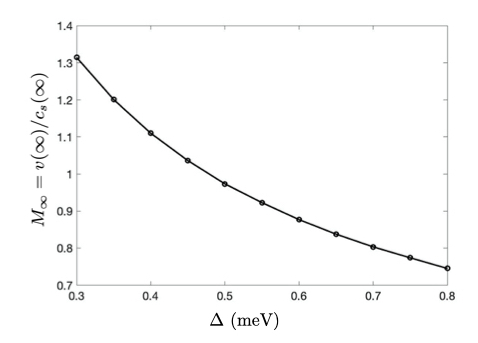}
\par\end{centering}
\caption{Mach Number $M=v_\infty/\sqrt{g n_\infty}$ vs $\Delta$ for $F=F_c$. Physical parameters:   $k_{p}=0.718 \;\mu\text{m}^{-1}$,
$m=1$~meV.ps$^2.\mu\text{m}^{-2}$,  $\gamma=0.05$ meV, $  g=0.01$ meV. We point out that   $v_\infty=\hbar k_p$ is a constant in the simulations with $\hbar=0.654$ meV.ps.}\label{fig:mach}
\end{figure}

We point out that in our case the Mach number is large (from 0.75 to 1.3), therefore the approximation
 $v_{max}\sim v_\infty (2+7/6*M^2)$ of Ref. \cite{rica} is not correct. Nevertheless, we have found that after the transition of emission of vortices, $v_{max}/v_\infty $ is not too far from 2.5 but we do not know how general this is and whether it can be estimated. Maybe the techniques in \cite{PNB} could help. We have plotted in Figure \ref{fig:mach} the Mach number at infinity vs $\Delta$, to show that it varies though the velocity at infinity is kept constant in our simulations. This can also be estimated from Eqaution \eqref{fn}.



 \subsection{Theoretical estimates}

Now we want to estimate $F_c$ analytically. As in \cite{FPR}, the operator in Equation \eqref{eq:cont} changes type when \begin{equation}
\partial_{v}(n(v)v)=0\label{eqtrans}
\end{equation} where $v=|\hbar\nabla \theta|^2$. In our case, from Equation \eqref{eq:ber}, the equation providing $n(v)$ is not simply quadratic in $v$ but is given as a polynomial of order 3 in $\sqrt{n(v)}$:
\begin{equation} gn(v)+\frac{v^2}2-\Delta+\frac{F}{\sqrt {n(v)}}\cos(\phi)=0.\label{Fn}\end{equation}
Instead of solving it, we prefer to differentiate it with respect to $v$ to find
\begin{equation}\left(g-\frac F{2\sqrt {n(v)} n(v)}\cos(\phi)\right)n'(v)+v=0.\label{n'}\end{equation} We identify $n'(v)$ from Equation \eqref{n'} and plug this into Equation \eqref{eqtrans} to find that at the point of maximal velocity, the cosine is equal to $-1$ and \begin{equation}v_{max}^2-gn -\frac{F}{2\sqrt n}=0. \label{vmaxcrit}\end{equation}This  changes sign at the transition where vortices are emitted. This is quite consistent with our numerical computations because indeed the critical value of $F_c$ corresponds to the case where this function changes sign. For instance, for $\Delta=0.3$, the function is negative for $ F=0.2515$ and positive for $ F=0.2712$  coinciding with the change of behavior reported in Figure \ref{fig:density}. We point out that Equation \eqref{vmaxcrit} is consistent with Ref. \cite{Nardin2011} where  the transition takes place when the Mach number $v_{max}/\sqrt{gn}$ is one. In our case, Equation \eqref{vmaxcrit} yields $v_{max}/\sqrt{gn}=\sqrt{1+F/2gn\sqrt n}$ which is close to 1, and equal to 1 when the driving is zero as in \cite{Nardin2011}.

\begin{figure}\begin{centering}
\includegraphics[width=.7\columnwidth]{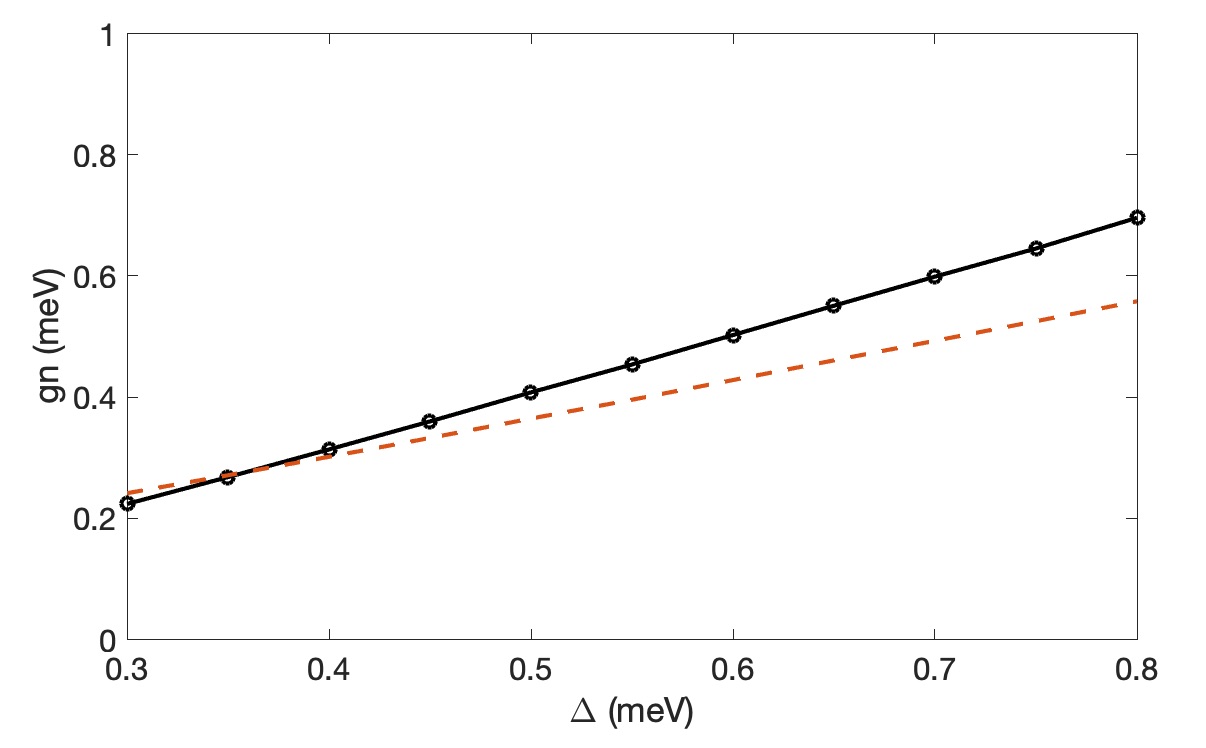}\par\end{centering}
\caption{$gn_{max}$ vs $\Delta$ computed from the numerics (thick line) and  coming from Equation (\ref{eqcrit}) (dashed line).}\label{fig4}
\end{figure}

  Because it is not easy to test Equation \eqref{vmaxcrit} numerically for many values, we  replace $v_{max}^2$  in Equation \eqref{vmaxcrit} from Equation \eqref{Fn} and we find an equivalent condition at the critical case:
  \begin{equation}\frac{gn_{max}}2-\frac{\Delta}3-\frac{F_c}{4\sqrt { n_{max}}}=0\label{eqcrit}\end{equation} This expression changes sign at the transition where vortices are emitted. This is verified by our numerical simulations leading to an error in Equation \eqref{eqcrit} of order $10^{-3}$ to $10^{-2}$. In Figure \ref{fig4}, we plot $gn_{max}$ from the numerics and compare it with the formula coming from Equation \eqref{eqcrit}, namely $2\Delta /3+F_c/2\sqrt {n_{max}}$. In fact, in this equation, the term in $F_c/\sqrt n$ is of lower order leading to an approximation
  \begin{equation}gn_{max}\sim \frac{2\Delta}3\label{nd}\end{equation} which is quite consistent with our numerics and relates therefore the Mach number to $\Delta$. Nevertheless, we notice a slight discrepancy between the numerics and the analytics which is likely to be due to errors in numerical estimates of $n_{max}$.

We hope that these estimates can help future experiments to determine the critical pump amplitude or the regime of parameters of interest.


\section{Conclusion}

We analyze the driven-dissipative nature of a polariton superfluid, in particular the effect of the pump amplitude and pump detuning. As they are varied, the solution goes from a superfluid solution to a solution emitting vortices. We can characterize analytically the change of behaviour and onset of turbulence by  Equation \eqref{vmaxcrit} or \eqref{eqcrit}. The relation between the parameters that we derive is consistent with the numerical simulations. In particular, we find that, if the driving field is kept fixed, as the pump detuning varies, the maximal density is proportional to the pump detuning at the critical transition.

\bibliographystyle{naturemag}
\bibliography{NotePourAmandine}

\end{document}